\documentclass[12pt]{article}
\usepackage{varioref,exscale,latexsym,amsmath,amssymb}
\usepackage{graphicx}
\def\beq{\begin{equation}}
\def\eeq{\end{equation}}
\def\beqa{\begin{eqnarray}}
\def\eeqa{\end{eqnarray}}
\begin{document}
\sloppy
\title{{\bf Dynamics of M-theory vacua}}
\author{John F. Donoghue\\ \\
Department of Physics\\ University of Massachusetts, Amherst, MA  01003 USA\\  \\
and \\ \\
Institut des Hautes \'{E}tudes Scientifiques \\
Bures Sur Yvette, 91440 France }
\begin{titlepage}
\maketitle
\begin{abstract}
At very early times, the universe was not in a vacuum state. Under the assumtion
that the deviation from equillibrium was large, in particular that it is higher than
the scale of inflation, we analyse the conditions for local transitions between
states that are related to different vacua. All pathways lead to an attractor
solution of a description of the universe by eternal inflation with domains that
have different low energy parameters. The generic case favors transitions between
states that have significantly different parameters rather than jumps between nearby
states in parameter space. I argue that the strong CP problem presents a potential
difficulty for this picture, more difficult than the hierarchy problem or the
cosmological constant problem. Finally, I describe how the spectrum of quark masses
may be a probe of the early dynamics of vacuum states. As an example, by
specializing to the case of intersecting braneworld models, I show that the observed
mass spectrum, which is approximately scale invariant, corresponds to a flat
distribution in the intersection area of the branes, with a maximum area $A_{\rm
max}\sim 100 \alpha'$.
\end{abstract}
\vspace{0.5 in}
 \setcounter{page}{0}
\end{titlepage}

\section{Introduction}

It is an empirical fact, yet a great puzzle, that the universe began out of
equilibrium. While we expect that every theory has at least one vacuum state, the
universe did not make use of such a state. We remain out of equilibrium even today.
Present indications suggest that we are approaching a state consisting of de Sitter
space with a vacuum energy of magnitude $\Lambda_0 = 2.7 \times 10^{-59}$~TeV$^4$,
although even this may not be the final resting state of the universe. In any case,
it is clear from observational evidence that the Universe is not now in a vacuum
state, and was not in the past.

Moreover, it is clear that the universe was further from equilibrium in the past. We
can reliably trace back the universe to a period of higher temperatures and faster
expansion. The ultimate scale describing the departure from equilibrium is less
clear. If our 4d world emerged from string theory, it is reasonable to assume that
this scale was the the mass scale of the 4d effective field theory - $E_4$ - which
we will assume is a fraction of the the string scale.

Finally, from observational evidence it appears increasingly likely that scalar
field inflation occurred in the early universe. The isotropy of the universe, the
overall density with $\Omega=1$ and the detailed pattern of density fluctuations in
the microwave background support this conclusion.  This has the implication that our
present observable universe emerged from a very small patch of the original universe
and remains only a small fraction of the full universe.

These ingredients lead to the expectation that in string theory the full universe
should consist of regions which involve different vacuum states. As analyzed in more
detail below, the dynamics of transitions between different ground states which
occur when the universe is far from equilibrium will lead to different domains with
different cosmological constants and other parameters. Rather than having a single
ground state that permeates the whole universe, as we tend to assume for most field
theories, the lack of equilibrium and the existence of inflation coupled with string
theory transitions will lead to a multiple-domain universe.

String theory has very many different vacua which are possible~\cite{douglas,
Ashok}. There are also mechanisms for transitions between them. A simple and well
known example of this is the membrane nucleation process initially studied by Brown
and Teitelboim~\cite{Brown1,Brown2}. In this situation, bubbles of a new vacuum form
in four dimensions with a different value of the effective cosmological constant
through the nucleation of a two dimensional membrane coupled to a four-form field.
At present energies, the probability of this nucleation is so small that it is of
little cosmological interest. However, in the early universe with inflation there
would be regions in which membranes are formed. Moreover, at high energies the
bubble formation need not involve tunnelling and I will present estimates for
nucleation in a finite temperature state. For a universe that is out of equilibrium
by $E_4$, these estimates suggest that causally disconnected regions will be related
to different ground states. In the subsequent cosmological evolution, inflation
would have placed these other regions outside of our observable horizon.

This paper is an exploration of the dynamics and phenomenology of such a universe.
The different regions of the present universe would potentially involve different
low energy theories or at least different parameters. A serious recent estimate
suggests that in there could be very many string vacua (perhaps $10^{100}$) that
look like that Standard Model with couplings and masses such as are observed, as
well as far more with different parameters~\cite{douglas,Ashok, Susskind, Dine1}.
This would make it unlikely that the prediction of these parameters would be a test
of string theory. I will make a preliminary exploration of a different notion -
namely that the quark and lepton masses provide a visible remnant of the early
dynamics, reflecting the weight or measure by which the Yukawa couplings are
distributed.

\section{The distance between vacuum states}

The low energy effective field theory is a mapping from string theory parameters
(manifolds, moduli, fluxes) to Standard Model parameters (masses and coupling
constants). The number of choices for compact manifolds, for embedded fluxes and for
branes wrapped on cycles is extremely large. Because of this large number, the
possible output parameters of the low energy theory are also quite large and quite
possibly are densely packed. Douglas\cite{douglas} has initiated a program of
counting the numbers of string theory vacua. Each vacuum state is a delta function
in string theory parameter space, since the fluxes and hence moduli are
quantized\cite{Polchinski, Bousso}\footnote{This neglects the possibility of
unquantized fluxes that can occur in warped universes with infinite
dimensions\cite{Kallosh}, in which case the number of vacua is infinite and
continuous.}. However if the resulting moduli potentials are closely packed, it
makes sense to define an approximate measure for the vacuum states
\begin{equation}
d\mu(V) = \sum_{T\in theories} \delta(V-V(T))
\end{equation}
The low energy vacua can appear almost as a continuum if the resulting parameters
are dense.  The estimates of the total number of possible flux vacua is difficult to
estimate precisely but could be of order $\sim 10^{300}$\cite{douglas}. Restrictions
to those parameters that match the Standard Model was estimated to reduce the number
by a factor of $\sim 10^{-140}$. However, this leaves the possibility that there
could be even order $10^{100}$ vacua that reproduce the parameters of the Standard
Model {\em within their present experimental error bars}. However, even if this high
degeneracy is not realized and our set of Standard Model parameters is somewhat
unique, in the sense that there is only a single string vacua that produces these
parameters within the present experimental error bars, the important result is that
there are an extremely large number of related vacua that differ by the change of
some flux or brane wrapping.

An essential point for the dynamics of these vacua is that two vacua that are close
together in Standard Model parameter space are far apart in the parameters of string
theory. Likewise the states that are close in string theory parameter space are
relatively far apart in their Standard Model parameters. The density of possible
Standard Model parameters occurs only because of a great multiplicity of different
fluxes vacuum choices. This feature is displayed in the work of Bousso and
Polchinski\cite{Bousso}. For the compactification of M theory on a seven-manifold
with several nontrivial three cycles, the quantization rules yields a vacuum energy
\begin{equation}
\Lambda = \Lambda_{\rm bare} + \sum_i n_i^2 \frac{\pi M_{11}^3 V_{3,i}^2}{V_7}
\end{equation}
where $n_i$ are integers and the $V_i$ are the invariant volumes. The spacing of the
allowed values of $\Lambda$ can be very small even if the scale of each contribution
is itself large. Bousso and Polchinski~\cite{Bousso} show that if there were of
order 100 such fields, the density of states would be so high that the differences
between two values of the cosmological constant would be of the order of the
experimental value. However, a change of a single flux by one unit would then change
the vacuum energy by a factor
\begin{equation}
\Delta \Lambda = \frac{2\pi n_i M_{11}^3 V_{3,i}^2}{V_7}
\end{equation}
Unless one of the three cycles is exceptionally small, this is a large jump of order
the compactification scales. In order to move between the two small values for the
cosmological constant that are each of the same size as the experimental value, one
would have to rearrange of order 100 different four-form flux values.

Changes in the internal fluxes also lead to changes in the moduli. Generically,
non-zero fluxes contribute to the potentials for the moduli fields
\begin{equation}
V\sim \sum_i F^2_i e^{-a_i \phi}
\end{equation}
(for example, see Ref.~\cite{Ovrut} for a discussion of these potentials in a
cosmological context.). A transition in the fluxes will readjust the minimum of the
moduli potentials leading to changes in the parameters of the low energy Standard
Model. The change of one flux by one unit will then also the modify the moduli by a
significant amount.

This behavior has a consequence for the dynamics of these fields in cosmology.
Simple changes in the string theory fluxes do not move between closely related
Standard Model vacua with small changes in the parameters. Instead the most likely
changes are far apart in the parameters of the Standard Model.

In the work of Feng et. al. \cite{wilczek}, an attempt was made to construct string
vacua where the cosmological constant had small steps - of the order of the
experimental value - when a four form flux was changed by one unit. This would allow
the relaxation of the cosmological constant via membrane nucleation, as described in
the next section, to proceed by small steps with an end result that could naturally
be of order the observed value. However this was accomplished only by going to
extremes of string theory parameter spaces, i.e. cycles of vanishing size or tiny
string couplings. A more generic vacuum state has larger steps. Moreover, even if
there are degenerating cycles, there can be other cycles of normal size in the
internal manifold that generate flux potentials. Then there will be additional flux
changes that are possible that lead to large jumps in parameter space. In this class
of theories there would then be both large and small jumps in the low energy
parameters upon membrane nucleation.

\section{Dynamical transitions}

In an evolving universe there can be transitions between different vacuum states. We
will consider two types of transitions. In de Sitter phase, the membrane nucleation
of Brown and Teitelboim will be relevant. However, in other circumstances finite
temperature nucleation will be relevant. These have not been discussed previously in
the literature and we will discuss these in more detail.

\subsection{de Sitter transitions}

Brown and Teitelboim calculated the probability that a 2d membrane will be nucleated
in 4d de Sitter spacetime, with the interior region having a different value of a
four-form field strength. The Brown-Teitelboim results are simply presented in the
notation of Ref \cite{wilczek} which summarize the formulas for the transitions more
compactly than those of the original work.

Four-form field strengths describe fields for whom the local equation of motion
require that they be constants in four dimensions The gauge potential gauge
potential is a three-form with a four-form field strength tensor \beq
F_{\alpha\beta\gamma\delta} =
\partial_{[\alpha} A_{\beta\gamma\delta ]} \eeq where the square brackets denote the
antisymmetrization of the indices.  The action \beq S_F =-{1 \over 48} \int d^4x
{\sqrt{-g}} ~F_{\alpha\beta\gamma\delta}F^{\alpha\beta\gamma\delta} \eeq leads to
the equation of motion \beq
\partial^\alpha \left[ \sqrt{g} F_{\alpha\beta\gamma\delta}\right] = 0    .
\eeq The only solution to this is \beq F_{\alpha\beta\gamma\delta} = {c \over
\sqrt{g}} \epsilon_{\alpha\beta\gamma\delta} \eeq for arbitrary constant $c$. Thus
this field is nondynamical, with only a constant solution. Substitution of this
solution in Einstein's equations
 shows that it behaves as a
positive cosmological constant.

Four-form fields are ubiquitous in string theory. Each four-form couples to a
2-brane source, generically labeled $B^\mu$. The action for the coupling of the
2-brane to the gauge potential $A_{\alpha \beta\gamma}$ is
\begin{equation}
S= \int d^3 x \left[ \tau_0 \sqrt{det (g_{\alpha\beta} ~\partial_a B^\alpha
\partial_b B^\beta )}+ \frac{\rho_0}{6} \epsilon^{a b c}\partial_a B^\alpha
\partial_b B^\beta
\partial _c B^\gamma A_{\alpha \beta\gamma}\right]
\end{equation}
Here $\rho_0$ is the charge per unit area of the 2-brane and $\tau_0$ is the tension
of the 2-brane. For BPS 2-branes the tension and the charge are related by
\begin{equation}
\tau_0 = \frac{\rho_0 M_P}{\sqrt{2}}
\end{equation}

In the process of membrane creation, the four-form potential will differ across the
membrane. Let us define the interior and exterior values of the four-form by
\begin{equation}
F^{\alpha\beta\gamma\delta}_{i,e} =
\frac{c_{i,e}}{\sqrt{g}}\epsilon^{\alpha\beta\gamma\delta}
\end{equation}
The overall cosmological constant will then have interior and exterior values
\begin{equation}
\Lambda_{i,e} = \Lambda_0 + \frac12 c^2_{i,e}
\end{equation}
Here $\Lambda_0$ is the cosmological constant from all other sources. We will use
the notation
\begin{equation}
\lambda = 8 \pi G \Lambda
\end{equation}
to differentiate the cosmological constant from the vacuum energy density. Note that
the former has units $E^2$ while the latter has units $E^4$.

When there is a transition to a lower value of the cosmological constant, BT show
that the bubble size that minimizes the instanton action is equivalent to the
condition of energy conservation in membrane formation. Specifically, for de Sitter
metrics of the form
\begin{equation}
ds^2 = -(1 - \frac13 \lambda r^2) dt^2 + (1 - \frac13 \lambda r^2)^{-1} dr^2 +r^2
d\Omega
\end{equation}
the energy increase associated with bubble nucleation includes both the vacuum
energies and the mass of the membrane,
\begin{equation}
\Delta E(r) = \frac43 \pi (\Lambda_{int} -\Lambda_{ext}) r^3 + 2\pi \tau r^2 \left[
\sqrt{1 - \frac{8\pi G \Lambda_{int}r^2}{3}} + \sqrt{1 - \frac{8\pi G
\Lambda_{ext}r^2}{3}} \right]
\end{equation}
The condition $\Delta E (r=b) = 0 $ fixes the radius of the membrane at nucleation
\begin{equation}
b = \left[ \frac{9 \tau^2}{(6\pi G \tau^2)^2} + 12 \pi G \tau^2
(\Lambda_{ext}+\Lambda_{int}) + (\Lambda_{ext}-\Lambda_{int})^2\right]^\frac12
\end{equation}
Although this derivation only holds for the case $\Lambda_{int} <\Lambda_{ext}$, the
same result holds for the situation where the cosmological constant increases in the
interior of the bubble, $\Lambda_{int} > \Lambda_{ext}$.

The rate of nucleation per unit volume has the form \beq \frac{\Gamma}{V} \sim
e^{-B} \eeq where B is the instanton action
\begin{equation}
B = \frac{3M_P^2}{16}\left[\frac{1}{\Lambda_{ext}}(1+ \cos
\theta_{ext})-\frac{1}{\Lambda_{int}}(1- \cos \theta_{int})  \right]
\end{equation}
where the angle factors are given by
\begin{eqnarray}
\cos \theta_{int} &=& \frac{\Lambda_{ext} - \Lambda_{int} + 6 \pi G
\tau^2}{[\Lambda_{ext} - \Lambda_{int} + 6 \pi G \tau^2]^2 + 24 \pi G \Lambda_{int}}
\nonumber \\
 \cos \theta_{ext} &=& \frac{\Lambda_{int} - \Lambda_{ext} + 6 \pi G
\tau^2}{[\Lambda_{int} - \Lambda_{ext} + 6 \pi G \tau^2]^2 + 24 \pi G \Lambda_{ext}}
\end{eqnarray}
Despite the overall factor of the Planck mass in $B$, the rate is independent of the
Planck mass when the other scales are much smaller.  In the limit, $G\to 0$ one has
\begin{eqnarray}
B &=& \frac{27 \pi^2}{2}\frac{\tau^4}{(\Lambda_{ext} - \Lambda_{int})^3}
~~~~~(\Lambda_{int} < \Lambda_{ext}) \nonumber \\
B &=& \infty ~~~~~~~~~~~~~~~(\Lambda_{int} > \Lambda_{ext})
\end{eqnarray}
This indicates that transitions which decrease $\Lambda$ can take place without the
mediation of gravity while those that increase the cosmological constant only occur
as a gravitational effect. A few other limits are also worth displaying. In the
limit of small $\tau$, i.e. $6 \pi G \tau^2 \ll |\Lambda_{ext} - \Lambda_{int}|$,
one has the related limit
\begin{eqnarray}
B &=& \frac{27 \pi^2}{2}\frac{\tau^4}{(\Lambda_{ext} - \Lambda_{int})^3}
~~~~~(\Lambda_{int} < \Lambda_{ext}) \nonumber \\
B &=& \frac{3}{8}\left[\frac{M_P^4}{\Lambda_{ext}} -\frac{M_P^4}{\Lambda_{int}}
\right] ~~~~~~(\Lambda_{int}
> \Lambda_{ext})
\end{eqnarray}
In the limit of small $\Lambda_{ext}$, ( $\Lambda_{ext} \ll 6 \pi G \tau^2 $) one
finds
\begin{equation}
B =\frac{3}{8}\frac{M_P^4}{\Lambda_{ext}}
\end{equation}
independent of $\Lambda_{int}$, while if $\Lambda_{int}$  is small ( $\Lambda_{int}
\ll 6 \pi G \tau^2 , \Lambda_{ext} $) one has
\begin{equation}
B =\frac{3M_P^2}{8}\frac{(( 6\pi G \tau^2)^2)}{\Lambda_{ext}(\Lambda_{ext} + 6\pi G
\tau^2)^2}
\end{equation}

\subsection{Finite temperature transitions}

At finite temperature, transitions need not take place by tunnelling, but can occur
through thermal excitation over the barrier\footnote{If the membrane instanton is
the analogue of the Schwinger mechanism producing an $e^+ e^-$ pair in a background
electric field, the thermal excitation discussed in this section is the analogue of
a thermal fluctuation producing an electron positron pair. }. The probability for
such transitions is given by the Boltzmann factor to reach the peak of the energy
barrier, namely
\begin{equation}
P \sim e^{-\beta E_*}
\end{equation}
where $\beta = 1/kT$ and $E_*$ is the height of the barrier\cite{langer, Linde1,
Linde2, Affleck, Gross, Garriga}.

In the energy equation, Eq 15, we may safely specialize to the case where
$\lambda_{ext} r^2 = 8\pi G \Lambda_{ext} r^2 $ is small compared to unity. This is
required to be small if the universe is to undergo finite temperature evolution -
otherwise it will quickly turn into de Sitter evolution. It is only if the exterior
cosmological constant is small that the finite temperature evolution will occur.
However, we should allow the possibility that the interior cosmological constant is
not small. Thus we will use
\begin{equation}
E(r) = \frac43 \pi (\Lambda_{int} -\Lambda_{ext}) r^3 + 2\pi \tau r^2 \left[ 1+
\sqrt{1 - \frac{8\pi G \Lambda_{int}r^2}{3}}  \right]
\end{equation}
The radius that corresponds to the peak of the energy barrier is obtained from the
condition \beq \frac{d E}{d r} = 0 \eeq

First consider transitions where the value of the cosmological constant decreases.
Since we already have $G\Lambda_{ext}$ small, we also will have $G\Lambda_{int}$
small. The energy is maximum at a radius \beq r_* =
\frac{2\tau}{\Lambda_{ext}-\Lambda_{int}}\eeq and will have the
value
\begin{equation} E_* = \frac{16\pi}{3}\frac{\tau^3}{(\Lambda_{ext}
-\Lambda_{int})^2}
\end{equation}
Note that this greatly favors transitions which lead to large changes in the
cosmological constant.

Because of the gravitational expansion, one can also have transitions to a higher
value of the cosmological constant. These occur through a thermal fluctuation which
is large enough that the interior expands. The tension would make bubble contract.
However if the interior is large enough the expansion of the interior can win. In
analyzing this situation, let us consider the case where $\Lambda_{int} \gg
\Lambda_{ext}$. The energy function can then be rescaled using the variable \beq x^2
= \frac{8\pi G \Lambda_{int}}{3}r^2 \eeq
\begin{eqnarray}
E(r) &=& \frac43 \pi \Lambda_{int} r^3 + 2\pi \tau r^2 \left[ 1+ \sqrt{1 -
\frac{8\pi
G \Lambda_{int}r^2}{3}}  \right]  \nonumber\\
&=& \frac{3\tau}{4 G \Lambda_{int}}\left[ \eta x^3 + x^2 (1+\sqrt{1-x^2})\right]
\end{eqnarray}
where
\begin{equation}
\eta = \sqrt{\frac{\Lambda_{int}}{6\pi G \tau^2}}
\end{equation}
The equation for the maximum of $E(r)$ is
\begin{equation}
\frac{dE}{dx} =0= 3\eta x +2 (1 + \sqrt{1-x^2}) -\frac{x^2}{\sqrt{1-x^2}}
\end{equation}
Since $\eta$ is positive, this will have a solution only due to the third term,
which can be traced back to the de Sitter metric factor in the energy. Hence we know
that this solution only exists due to the action of gravity. The interesting
limiting cases correspond to energies at maximum of
\begin{eqnarray}
E_* &=& \frac{8\tau}{9G\Lambda_{int}}  ~~~~~~~~ \eta \ll 1 \nonumber \\
&=& \sqrt{\frac{3}{32 \pi G^3 \Lambda}_{int}}~~~~~~~~ \eta~ {\rm large}
\end{eqnarray}
It is not hard to solve for the energy numerically at any given value of eta, but
the form
\begin{equation}
E_* =\frac{8\tau}{9G\Lambda_{int}} [1+ \frac98 \sqrt{\frac{3\Lambda_{int}}{32\pi G
\tau^2 }} ]
\end{equation}
provides an excellent interpolating formula for intermediate values of $\eta$. In
each case, transitions to the largest values of $\Lambda_{int}$ are favored.

\section{Multi-domain eternal inflation}

Let us follow the logic dictated by the assumptions described above. In this
section, we will systematically explore the possible evolutionary pathways of an
energetic early universe. The major ingredients are the assumption that
scalar-field inflation has taken place in the past evolution of our domain and that
the initial state of the universe was at an energy scale, called $E_4$ below, which
was greater than the scale associated with scalar-field inflation.

\subsubsection{Types of eternal inflation}

Essentially all of the pathways will involve at least one inflationary epoch. There
are two types of inflationary expansion that are worth distinguishing here 1)
Inflation driven by the energy density associated with a scalar
field~\cite{inflation} and 2) Inflation dominated by the energy associated with a
four-form field. In reality both the energy density of the scalars and the
four-forms contribute to the single parameter - the cosmological constant - that
drives the de Sitter phase~\cite{Mtheoryinflation}. However, the important
distinction comes in how the inflationary phase ends or changes. With a scalar
field, the field eventually will roll down a potential and dump all of its energy
which is converted into particles in the process of reheating. This leads to the
inflationary prediction that $\Omega =1$. On the other hand a four-form makes a
transition to a different value through the process of membrane nucleation. The
membrane will carry off some of the energy as the bubble expands, leaving the
interior region with less than the critical density, $\Omega<1$. For this reason,
four-form inflation is not a good candidate for the final inflationary phase which
appears to have taken place in our portion of the universe~\cite{Bousso2}.

It is important to keep in mind that both forms of inflation are generically future
eternal. In the case of scalar field inflation this is well known\cite{Vilenkin,
Steinhardt, Lindeeternal, Guth}. The scalar field that is responsible for inflation
will in some place fluctuate higher up the potential - in others it will fluctuate
down . While inflation will end in some regions as the field moves down the
potential, there will always be other domains that continue to inflate. In the case
of four-form inflation the reasoning is different. The membrane nucleation process
that changes the cosmological constant only does so on the interior of a finite
region. The exterior region continues to inflate at the old value of the
cosmological constant. Because the bubble expands at the speed of light while the de
Sitter expansion is more rapid, the bubbles do not fill the space and there are
always the exterior regions that are still inflating\footnote{This pattern also
occurred in ``old inflation'' in which there was a first order phase transition. }.
Thus once the universe enters an inflationary phase, it will always be inflationary
except in domains in which transitions to small values of the cosmological constant
has taken place.

\subsection{deSitter dominated}

We have assumed that the initial state of the universe is out of equilibrium by an
amount labelled $E_4$.  There will be various evolutionary pathways possible. An
immediate distinction is whether this energy is manifest as a cosmological constant
of order $\lambda = G \Lambda \sim G E^4_4$ or if it appears in the form of
energetic particles. In the former case the universe will immediately be in a de
Sitter phase (a ``cold'' start), while in the latter we will argue that it
thermalizes (a ``hot'' start) . The pathways will then branch out from these cases
depending on the nature of the subsequent transitions.

\subsubsection{From deSitter to de Sitter}

If vacuum energy is of order $\Lambda \sim E_4^4$, then the cosmological constant is
\begin{equation}
\lambda = 8 \pi G \Lambda \sim \frac{E_4^4}{M_P^2}
\end{equation}
The initial state is then one of de Sitter expansion with the dominant ingredient
being the cosmological constant from the four-form field.

In a de Sitter phase, the Brown-Teitelboim nucleation will be operational. Bubbles
of different values of the cosmological constant will be formed in the overall de
Sitter phase. Because the inflation is future eternal, there will be an ever
increasing number of the bubbles formed even if the probability for any one bubble
is not large. These transitions can be either to smaller or to larger values of the
cosmological constant. While transitions to larger values are possible in the
Brown-Teitleboim calculation, they may not be physically possible if they correspond
to such a large value that the low energy effective theory is no longer applicable.
However, if the effective theory is possible, then inflation simply continues and
further membrane nucleation takes place.

The regions that are finally most interesting to us are those that make a transition
to a smaller value of the cosmological constant. In these regions other behaviors
are possible. In some regions, there will be a transition to a lower cosmological
constant. Let us first neglect the possibility of particle production during this
transition. In this case, the patch of the universe will transition to another de
Sitter region. If the cosmological constant in this region is dominated by the
four-form field, we will repeat the above set of choices. If it is dominated by the
scalar field, we enter a period of scalar field inflation. The subsequent evolution
of such a domain is standard for the inflationary literature. It involves an
inflationary period, a brief roll-down for the scalar field and a reheating period,
followed by thermal evolution of the universe. Some of these domains have the
potential to evolve into the universe that we see observationally.

\subsubsection{From deSitter to thermal to inflation}

Another possibility is that at the time of the Brown-Teitelboim transition, the
domain experiences significant particle production. This is not present in the
original Brown-Teitelboim calculation, but it would clearly be present in string
theory. This is because the form fields also contribute to the moduli potentials. A
change in the moduli potentials will lead to a modification of the masses and
couplings of the low energy theory. Anytime these parameters change abruptly there
will be particle production. For example, the moduli potential will experience a
shift of order
\begin{equation}
\Delta V \sim \Delta F^2 e^{-a\phi}
\end{equation}
Since this is a steep potential the field will rapidly seek its new minimum,
producing particles as it readjusts. The reheating temperature cannot be calculated
without knowledge of the other contributions to the moduli potentia but, since it is
scaled by the change in the vacuum energy with no extremely small parameters
involved, the reheating temperature should be a fraction of the original vacuum
energy. These domains will resemble an open universe with a density below critical
density.

In the presence of particle production, there will be some domains that will evolve
differently from the above description. In these domains the residual cosmological
constant is smaller than the reheating temperature. These will then evolve as a
radiation dominated universe rather than as a de Sitter state. However, since the
last transition is that of four-form inflation, these domains have a density less
than the critical density and they will evolve to an almost empty open universe
unless de Sitter expansion takes over again in their future. Such domains are not
like ours. We could not find ourselves in an empty domain so that we need not
consider this case further.

\subsection{Particle dominated initial conditions}

In this section, we consider the case where the cosmological constant is smaller
than other forms of energy contained in the fields. In the absence of the expansion
of the universe it is clear that any such energetic initial condition would
eventually lead to thermal equilibrium. In the presence of expansion one must ask if
the rate of expansion prevents the system from reaching an approximate
equilibrium\cite{weinbergbook}. Individual cross sections and particle densities at
an energy $E$ will be of order
\begin{equation}
\sigma \sim \frac{\alpha^2}{E^2} ~~~~~~~~~~~n \sim E^3
\end{equation}
where $\alpha$ is the coupling strength of the interactions. At high energies, the
gauge interactions are characterized by an interaction strength of order $\alpha\sim
1/25$. Reactions that lead to thermalization then occur at a rate
\begin{equation}
{\rm rate} \sim g_* \sigma n \sim g_* \alpha^2 E
\end{equation}
where $g_*$ describes the number of particles available, which could be of order
$10^2$. On the other hand, gravitational expansion could prevent equilibrium from
occurring if it is more rapid than the equilibration rate.  The expansion rate is
\begin{equation}
H \sim \sqrt{G\rho} \sim \sqrt{g_*GE^4}
\end{equation}
The ratio of these rates is
\begin{equation}
\frac{\sigma n}{H} \sim \frac{g_*^{1/2} \alpha^2}{\sqrt{G E^2}}
\end{equation}
We see that unless the energy scale is close to the Planck scale, the universe will
thermalize. In the case that the universe starts out close to the Planck scale,
there will not initially be thermal equilibrium. However the expansion will scale
down the energy by a factor of the scale factor
\begin{equation}
E(t) \sim \frac{a(t_0)}{a(t)}E_0
\end{equation}
In this way the energy density will eventually fall to a value which does allow for
thermal equilibrium. Thus in cases where the cosmological constant does not play a
role in gravitational expansion, the gauge interactions will eventually lead to
thermal equilibrium.

In this situation, finite temperature transitions are initially important. As
described in Sec 3, this can result in thermal creation of regions of different
values of the cosmological constant. Transitions with a large change of the
cosmological constant are favored. Initially, the effect of the cosmological
constant is subdominant - it is hidden beneath the larger thermal energy. The
subsequent evolution of the various domains depends on the magnitude of the
cosmological constant generated by the thermal transitions.

\subsubsection{From thermal to deSitter}

Since the thermal transitions are exponentially suppressed, most regions of the
universe will evolve without any change in the value of the cosmological constant.
However, as the thermal region cools it will generically enter a period of
inflation. (The special case of no inflation is dealt with below.) Depending on the
magnitude of the different contributions to the cosmological constant, this can be
either scalar field inflation or four-form inflation. In either case, the
description of the universe becomes that of de Sitter expansion and the energy
contained in the field degrees of freedom rapidly goes to zero. If the four-form
fields dominate the cosmological constant, the situation reverts to the analysis of
the previous subsection. If it is the scalar field energy that dominates the
cosmological constant, then the usual scenario of eternal scalar field inflation
results. However, in addition to the fluctuations of the scalar field, one needs to
account for the Brown-Teitelboim fluctuations of the four-form field, also described
above. Hence in either of these cases, we arrive at a universe of eternal inflation
with domains that have fluctuated to different values of the cosmological constant.
A subset of these domains can appear similar to our own.

In some regions, transitions to other values of the cosmological constant will
occur. If the resulting cosmological constant remains smaller than the thermal
energy density, the analysis of the previous paragraph remains valid. If the final
cosmological constant is larger that the local thermal energy, then the universe is
locally de Sitter dominated. The subsequent history of these domains then follows
the pathways described above.

Finally, there will be regions that settle into a low energy state (with energy E
$<$ E(inflation)) without going through any form of inflation. For this to occur,
the total value of the cosmological constant needs to be small, so that both the
four-form contribution and the scalar field contribution are individually small.
Such regions can exist if the initial cosmological constant is tiny or if there was
a thermal fluctuation to a tiny value of the cosmological constant. These regions
are excluded from being our observable universe by the initial assumption that our
domain has gone through a period of scalar field inflation. Moreover, a region such
as this is an infinitesimal fraction of the allowed regions as long as a single
transition occurs to start the process of eternal inflation.

\subsection{All pathways lead to inflating domains}

Overall, this analysis has lead us to a universe that has regions of eternal
inflation with different values of the expansion rate. In this sense, such a
multi-domain inflating universe is an attractor solution. For all energetic initial
states the dynamical transitions which are possible in string theory will lead to
this as the final state. The different regions will span the various allowed values
of the physical parameters. In a subset of those regions, inflation will have ended
and a matter dominated era can occur. Some of these regions will be similar to our
own.

\section{Typicality and the strong CP problem}

Even if there are very many domains, this does not absolve us from needing to
understand the structure of fundamental theory in our domain. Physics is an
experimental science and we experimentally explore the nature of our domain.
However, the multiple domain structure does change the nature of some of the key
questions. Within the framework under discussion I will argue that this selects the
strong CP problem as a more important problem than the hierarchy problem or the
cosmological constant problem\footnote{While this paper was being written up, Ref.
\cite{Dine1}
 appeared
which, containing related comments on the strong CP problem and supersymmetry
breaking. }.

There are three fine-tuning problems that are usually highlighted as violations of
the principle of ``naturalness'' - the cosmological constant problem, the hierarchy
problem and the strong CP problem. The principle of naturalness states that it is
unnatural for a parameter in a theory to be much smaller than the magnitude of the
radiative corrections to that parameter. If a parameter is unnaturally small, then
there must be a fine-tuned adjustment of many large contributions to sum up to a
value much smaller than any of the individual contributions. Since this is
aesthetically distasteful, it motivates searches for new dynamical mechanisms in
which the smallness of the parameter is natural. The expectation that we will find
new dynamics at around 1 TeV is primarily motivated by the argument of naturalness
for the Higgs vacuum expectation value. Note that naturalness is a somewhat fuzzy
concept - witness the discussions of whether the present constraints on
supersymmetry make that theory unnatural. However, it is an effective motivator for
deciding which problems are important to study for indications of new physics.

In a theory with multiple domains, naturalness is not as useful a concept. In the
ensemble of domains there will always be some with unnaturally small values for the
parameters. In addition, whether we like it or not, one must unavoidably take into
account ``anthropic'' boundary conditions~\cite{anthropic}. Such constraints will
sometimes select only those domains which are technically unnatural. For example, we
could not conceivably find ourselves in a domain with natural values of the
cosmological constant, so we inevitably must constrain our consideration to domains
with a viable value of the cosmological constant.

In multiple domain theories, there is a different concept that can replace
naturalness. Within a given fundamental theory with a particular history, the
ensemble of domains will define a distribution for the parameters of the low energy
theory. Some of these parameters may be restricted if we specialize to the
anthropically allowed domains, yet others will still have a significant range and
distribution. We would expect that our domain should be typical of this ensemble,
subject to anthropic constraints. The parameters that we find should not be
extremely unusual for the ensemble of viable domains. I will refer to this
expectation as ``typicality'' \footnote{ Vilenkin~\cite{mediocrity} has provide a
more specific formulation of this idea under the name of the ``principle of
mediocrity''. Vilenkin's principle of mediocrity has a technical difference with my
usage of typicality. The former is defined by a measure which is proportional to the
number of civilizations in a given domain. While this emphasis on the number of
civilizations may be laudable in certain contexts, I prefer not to include it in the
present discussion. Moreover, mediocrity has the unfortunate connotation of ``not
good enough'', while we have a pretty damn good universe.} and will provide examples
of how it changes the motivation for new dynamics beyond the standard model in
multiple domain theories. In the next section, I will extend this notion to obtain
specific information on the initial ensemble of domains.

There are only a few parameters of the Standard Model which have significant
anthropic constraints. Primary among these are the cosmological constant and the
Higgs vacuum expectation value. Physically, these constraints are manifest in the
requirement that the universe allows matter to clump into stars- which strongly
restricts the values of the cosmological constant\cite{weinberg} - and in the need
for atomic elements beyond hydrogen to be stable - which only occurs for a narrow
range of the Higgs vev\cite{agrawal}. In a multiple domain realization of string
theory, we would restrict our attention to only those vacua that satisfy these
constraints. Thus the naturalness or fine-tuning problems of these two parameters
are not significant problems for such multiple domain theories.

However, the strong CP problem appears to be in severe conflict with typicality. In
the Standard Model the $\theta$ parameter, which measures the amount of strong CP
violation, is a dimensionless coupling constant. It is infinitely renormalized by
radiative corrections and there is no reason within the theory for it to be small.
There are also no known anthropic constraints on the value of $\theta$ - the world
would be essentially unchanged for $\theta$ of order unity. Nevertheless,
experimental bounds on the electric dipole moment of the neutron constrains \beq
\theta \le 10^{-10}  \ \ . \eeq This is a potential problem for multiple domain
theories. A priori, one would expect that the ensemble of viable Standard Model
domains would have a distribution of $\theta$ that would span values much larger
than the experimental constraint by up to ten orders of magnitude. If this is the
case, we would have to conclude that we are a very non-typical domain. Since it is
not very likely that we would randomly find ourselves in such a domain, we need to
seek dynamical solutions to the strong CP problem. Moreover it suggests that these
solutions must be generic in string theory - that they occur in a typical vacuum
solution. This is a strong constraint. In multiple domain theories, the value of
$\theta$ appears more puzzling than that of $\Lambda$ or $v$.

Typicality may also have other implications. As mentioned above, a small
cosmological constant and a low Higgs vev are required by anthropic constraints, so
we should only look for vacua that live within this range and ask if our vacua is
typical of this range. But this does not exhaust the issue of typicality. For
example, in principle one could decide if low energy SUSY is likely in string
theory. One would do this by counting the number of available string vacua. Are
there more viable SM vacua with low energy SUSY breaking or with high energy
breaking? These numbers are unlikely to be similar. As a hypothetical extreme
example, consider a situation where there are of order $10$ viable SM vacua with
Planck scale SUSY breaking and $10^{120}$ with weak scale breaking. The differences
in these numbers could arise because of the need to have an appropriately small
cosmological constant, which might be less likely if the SUSY scale is larger. In
this case, typicality predicts that we should find supersymmetry at low energy. This
would be a {\it statistical} prediction, but could be compelling - the numbers are
potentially just too overwhelming. Note also that this prediction could be different
from naturalness - the numbers of the different vacuum states could in principle
been reversed such that there are more viable vacua with high energy SUSY
breaking.\footnote{Michael Douglas (private communication) has also suggested that
this distinction could be indicative of the mechanism of SUSY breaking - perhaps
there are far more vacua with an appropriate $\Lambda$ with gauge mediated SUSY
breaking than gravity mediated breaking because of the lower mass scale associated
with the former.} The counting of these numbers of states is a well-posed question
that can in principle be addressed in string theory.

Dark matter is also a potential problem with the principle of typicality. Matter of
any form was inessential in early universe, yet in our present universe there are
comparable mass densities from ordinary matter and dark matter - modulo a factor of
5. If matter and dark matter come from different sectors of the fundamental theory
and correspond to quite different masses, why should their densities be so close in
the universe? It is not enough to find some values of the parameters that allows
such a relation - it must be a typical case. The generic solution to the nature of
dark matter should therefore likely have both dark matter and ordinary matter tied
to the same mechanism of production.

\section{Quark masses as a probe of vacuum dynamics}

If the parameters of the low energy theory are not unique predictions of string
theory, and we have only one domain to observe, it appears difficult to use the
parameters as probes of string theory.  However, there are many parameters in the
Yukawa sector and these may be subjected to a statistical analysis. We have 6 quark
masses, 3 lepton masses and the CKM parameters which are representative of the
distribution of Yukawa couplings. Even though this is not a very large number of
observables, we can still use them to obtain valuable information on the underlying
theory.

The basic idea is that if there are enough vacuum states consistent with the
Standard Model, the quark masses could appear as random variables distributed with
respect to some weight. Even if we held fixed one mass, there could be an ensemble
of domains that have a variety of mass values for the other quarks. The masses which
are observed would not be uniquely predicted but would be representative of this
ensemble. Observationally, the masses are not uniformly distributed - they are most
numerous at low mass, yet extend out to the very large top quark mass. As I will
discuss below, they are quite close to a {\em scale invariant} shape $\rho(m) \sim
1/m$. This weighting of the masses would then be the observational remnant of the
original ensemble.

In what follows, I will treat the quark masses as independent quantities. I will
also neglect any anthropic constraints, aside from some brief comments below.
Neutrino masses will not be considered here as they likely involve a different
mechanism from the other fermion masses.

The weight or measure of quark masses is defined as follows\cite{weight}. In an
ensemble of domains similar to our own, with the other Standard Model parameters
equal to ours, the fraction of masses found at a value $m$ within a range $dm$ is
defined to be
\begin{equation}
 f(m) = \rho (m) ~dm
\end{equation}
where $\rho (m)$ is the symbol for the weight. The normalization of the weight is
\begin{equation}
 1 = \int \rho (m) ~dm \  \ .
\end{equation}
The weight depends on the scale at which it is defined, and the renormalization
group equations for the weight were worked out in Ref. \cite{weight}. The comments
below apply to the weight at the weak scale.

One cannot simply plot the observed masses in order to reproduce the shape of
$\rho(m)$, because the masses would form a delta function distribution. One needs to
smooth this distribution in order to compare theory and experiment. One way to do
this involves taking the Hilbert transform
\begin{equation}
H(z) = \int^\infty_0 dm ~{z \rho (m) \over m + z}
\end{equation}
Here one can use the experimental masses to produce an $H_{\rm expt}(z)$, and
compare that form to the transform of various trial forms for $\rho(m)$ For the
experimental side we use
\begin{equation}
\rho_{exp} (m) = {1 \over N}\Sigma^N_{i=1} \delta(m - m_i)
\end{equation}
In \cite{weight}, the uncertainties associated with the experimental distribution
were assessed by variously dropping one of the quark masses from consideration (to
simulate the limited amount of statistics involved), by adding lepton masses either
raw or rescaled by a renormalization group factor, and by various ways of including
the CKM matrix elements. The experimental transform and its estimated uncertainty
are shown in Fig. 1.

\begin{figure}[h]
%\vspace*{5pt}
\begin{center}
\begin{minipage}[t]{.07\textwidth}
    \vspace{0pt}
    \centering
    \vspace*{35pt}
    \hspace*{-10pt}
    \rotatebox{0}{H(z)}
  \end{minipage}%
 \begin{minipage}[t]{0.93\textwidth}
    \vspace{-0pt}
    \hspace{-10pt}
    \centering
    \includegraphics[width=0.99\textwidth,height=!]{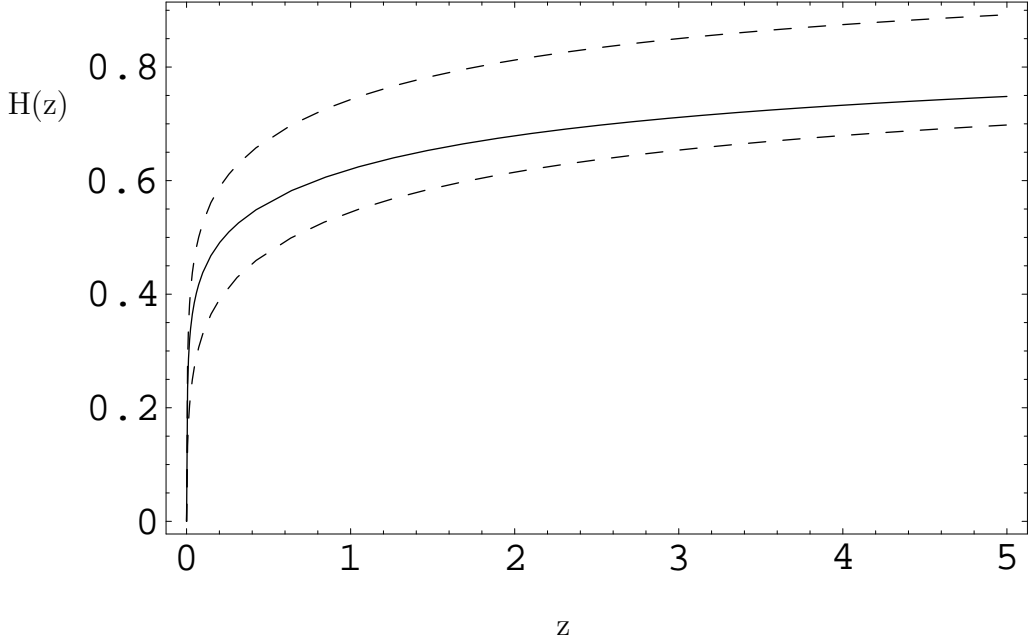}
    %${} \vspace*{-10pt}$
  \end{minipage}
\end{center}
\hspace*{0.52\textwidth} {z}\caption{The transformed weight H(z)corresponding to the
experimental values of the quark masses defined at the scale $M_W$ (solid curve).
The dashed curves are estimates~\cite{weight} of the upper and lower ends of the
allowed range of this quantity due to the limited statistics in the number of quark
masses. \label{weight} \vspace*{10pt}}
\end{figure}

In Ref \cite{weight} power-law weights were considered.  One form was a pure power
behavior combined with a cutoff at the quasi-fixed point of the renormalization
group $m^* \sim 220$~GeV
\begin{equation}
\rho_1(m) = {N \over m^\delta} \Theta( m^* - m)
\end{equation}
Here $\delta < 1$. The best fit was found for $\delta = 0.91$, and the result is
shown in Fig. 2. Also shown in this figure are the results for weights with lower
powers of $\delta$. One can see that the experimental range does significantly
constrain the form of the weight and that powers near unity are favored.  Because of
this fact , it is also useful to specifically consider a scale invariant form with
$\delta = 1$. Here we require a cutoff at low mass if the distribution is to be
normalizable.
 We can form a normalizable
 \begin{equation}
  \rho_3(m) = {N \over m} \theta (m - m_{\rm min}) \theta ( m^* - m)
\end{equation}
with $N = 1/ ln(m^*/m_{\rm min})$. The results depend only weakly (logarithmically)
on the lower cut-off. The result plotted in Fig 3 uses $m_{\rm min} = 0.1 m_e$. A
weight with $\rho (m) \sim 1/m$ can be described as ``scale invariant'' in the
following senses. In the first place, there is no scale in the shape of $\rho (m)$,
and the normalization constant is dimensionless and independent of the overall
scale. In addition, under any linear rescaling of the masses such as occurs for the
scale dependence of QCD,
\begin{equation}
m_2(\mu_2) = \left( {\alpha_s(\mu_2) \over \alpha_s(\mu_1)} \right)^{d_m}~
m_1(\mu_1)
\end{equation}
the renormalization group transformation rule\cite{weight} tells us that this weight
will remain unchanged (again, aside from the endpoints), since
\begin{eqnarray}
\rho_\mu (m) & = & \rho_{\mu_1}\left( m \left({\alpha_s(\mu_) \over \alpha_s(\mu_1)}
\right)^{-d_m}\right) \left({\alpha_s(\mu_)
\over \alpha_s(mu_1)}\right)^{-d_m}  \\
& = &  {1 \over  m \left({\alpha_s(\mu_) \over \alpha_s(\mu_1)} \right)^{-d_m}}
\left({\alpha_s(\mu_)
\over \alpha_s(\mu_1)}\right)^{-d_m} \\
& = & {1 \over m} .
\end{eqnarray}
It is intriguing that a scale invariant weight is close to the distribution seen by
experiment.

\begin{figure}[h]
%\vspace*{5pt}
\begin{center}
\begin{minipage}[t]{.07\textwidth}
    \vspace{0pt}
    \centering
    \vspace*{35pt}
    \hspace*{-10pt}
    \rotatebox{0}{H(z)}
  \end{minipage}%
 \begin{minipage}[t]{0.93\textwidth}
    \vspace{-0pt}
    \hspace{-10pt}
    \centering
    \includegraphics[width=0.99\textwidth,height=!]{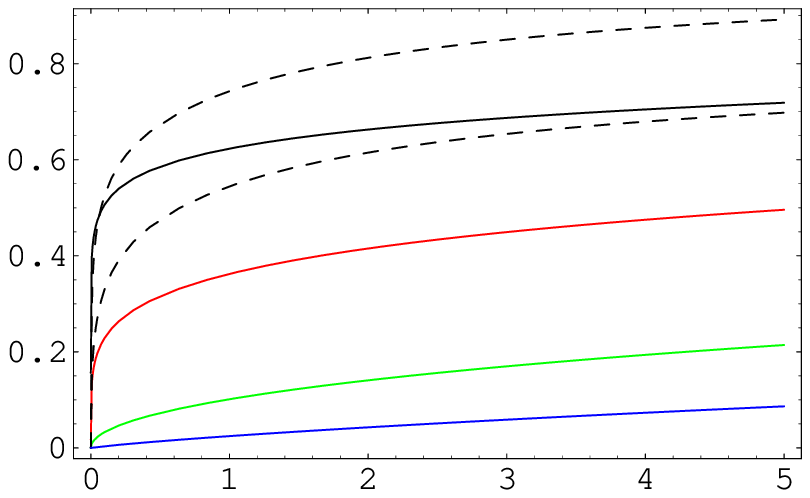}
    %${} \vspace*{-10pt}$
  \end{minipage}
\end{center}
\hspace*{0.52\textwidth} {z}\caption{The transformed weights H(z) corresponding to a
power law weight with exponent $\delta =0.91$ (black solid curve) compared to the
allowed experimental range (dashed curves). Also shown for comparison  below the
allowed range are power law weights with exponents $\delta =0.8$ (red), $\delta
=0.5$ (green) and a flat weight $\delta=0$ (blue), which shows that these weights
are experimentally disfavored. \label{powerweights} \vspace*{10pt}}
\end{figure}

\begin{figure}[h]
%\vspace*{5pt}
\begin{center}
\begin{minipage}[t]{.07\textwidth}
    \vspace{0pt}
    \centering
    \vspace*{35pt}
    \hspace*{-10pt}
    \rotatebox{0}{H(z)}
  \end{minipage}%
 \begin{minipage}[t]{0.93\textwidth}
    \vspace{-0pt}
    \hspace{-10pt}
    \centering
    \includegraphics[width=0.99\textwidth,height=!]{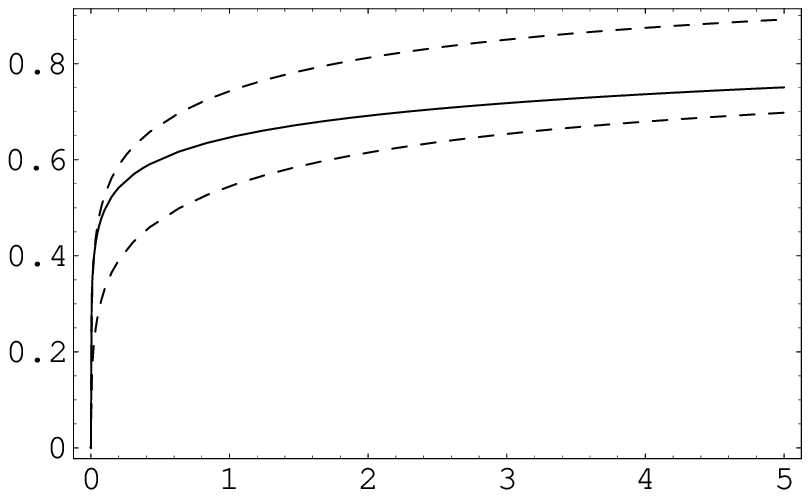}
    %${} \vspace*{-10pt}$
  \end{minipage}
\end{center}
\hspace*{0.52\textwidth} {z}\caption{The transformed weight H(z) corresponding to
the scale invariant weight $\rho \sim 1/m$ (solid curve), compared to the allowed
experimental range (dashed curves)  \label{scaleinvariant} \vspace*{10pt}}
\end{figure}

At this point let us specialize to the Intersecting Braneworld
models\cite{ibw,Ibanez1,Ibanez2,Ibanez3,Ibanez4}, because these have a concrete
realization of the physics underlying the Yukawa couplings of the Standard Model. In
these theories, chiral fermions live at the intersections of Dp branes in
compactified extra dimensions. The right-handed and left-handed fermion fields occur
at different intersections and the Higgs field exists at a third intersection. These
fields then do not have direct contact interactions, but the Yukawa couplings
connecting them through the world sheet instantons with an action proportional to
the area connecting the intersections. For a Yukawa coupling $h_{ijk}$  connecting
left-handed fermion $\psi_{Li}$, right-handed fermion $\phi_{Rj}$ and Higgs $H_k$
one has\cite{ibw}
\begin{equation}
h_{ijk} = h_0  e^{-\frac{A_{ijk}}{2\pi \alpha'}} e^{2\pi i \phi}
\end{equation}
where $A_{ijk}$ is the worldsheet area of the triangle connecting the three
intersection points of the three branes.

If the intersections relevant for the quark and lepton Yukawa couplings are
determined dynamically in the early universe, then the distribution that we see for
masses will reflect the distribution of brane intersections. What is interesting is
that a very simple distribution results. The scale invariant weight for quark masses
described above corresponds to a {\em flat} distribution of the worldsheet areas. To
see this we note that the exponential dependence in the area implies that
\begin{equation}
\frac{dm}{m}= \frac{d h}{h} = - \frac{dA}{2\pi\alpha'}
\end{equation}
Therefore if we define a weight or distribution for the areas via
\begin{equation}
\rho (m) dm \sim \rho(A) dA
\end{equation}
then the scale invariant weight $\rho(m) \sim 1/m$ implies a flat weight for the
area
\begin{equation}
\rho(A) = {\rm constant}
\end{equation}
A flat distribution of the areas is perhaps the most natural distribution.

There is also somewhat weaker information in the limits of the weight function if
the shape is exactly the scale invariant form. A flat distribution cannot extend to
arbitrarily large areas and still be normalizable. The upper range of the area
determines the ratio of the minimum mass to the maximum mass. The largest possible
values of the masses are obtained for areas close to zero, which of course is always
able to be realized. The minimum mass corresponds to the maximum area, via
\begin{equation}
h_{min} = h e^{-\frac{A_{max}}{2\pi\alpha'}}
\end{equation}
Thus in the ratio of masses, the overall scale drops out and one finds
\begin{equation}
\frac{m_{min}}{m_{max}} = e^{-\frac{A_{max}}{2\pi\alpha'}}
\end{equation}
If we take the minimum mass to be $m_{min} \sim 0.1 m_e$, we find
\begin{equation}
A_{max} \sim 2\pi\alpha' \ln (\frac{m_t}{0.1 m_e}) \sim 100 \alpha'
\end{equation}
Phenomenologically, there is only a weak constraint on the minimum mass as it enters
the weight only logarithmically. However, the minimum mass also only enters
logaritmically in the relation for the maximum area.

 One of
the outcomes of the intersecting braneworld models is then that a uniform
distribution in the internal parameters of the model, in this case the area,
translates to a very non-uniform distributions in the parameters of the low energy
effective theory. It is also interesting that the natural distribution in the model
goes a long way towards explaining the very puzzling distribution of fermion masses
seen in experiment - with an increasing density at low mass. What remains is to
better understand the dynamics that can produce the flat distribution of areas.

In this analysis, we have not considered the impact of anthropic constraints. As
discussed in \cite{weight} these have the possibility of distorting the experimental
weight. Further work would be required to determine the consequences of anthropic
constraints on this analysis.

\section{Final comments}

The initial assumptions are 1) an initial string theory state that is out of
equilibrium by an amount $E_4 > E_{\rm inflation}$, and 2) that our domain underwent
scalar-field inflation in the past. The primary physics ingredients have been 1) a
huge number of vacuum states as suggested in string theory, 2) the dynamical
mechanisms for transitions between them (tunnelling and thermal transitions). In
this situation, the universe that results will generically consist of multiple
domains related to different vacua.

Linde\cite{inflation} has emphasized how inflation creates the opportunity for
domains with different physical parameters, because an inflating universe leads to a
domain structure of regions inflating at different rates. However, this is not
sufficient. One also needs a mechanism for producing different couplings in the
different regions - otherwise all domains will eventually settle down to the same
vacuum state. String theory has such physical mechanisms.  The transitions between
different flux vacua -whether in de Sitter expansion or at finite temperature - will
necessarily produce differing parameters in some domains in an large and eternally
inflating universe.

Under the assumptions of this paper, the physics of our domain is seen to be
determined not by a condition selecting a unique ground state from string theory,
but from the particular past history of our patch of the universe. The attractor
solution that we have found consists of eternal inflation, with various domains in
which inflation has ended. Our domain is presumably one of these domains and is
clearly one which has satisfied certain anthropic boundary conditions. In such a
universe, it is unfortunately difficult to make unique predictions. However, we have
raised at least the hope of testing this picture through the use of statistical
predictions such as solution of various of the typicality problems, such as the
strong CP problem, and through the spectrum of fermion masses.

\section* {Acknowledgements}

I would like to thank Michael Douglas, David Kastor, Jussi Kalkkinen and Thibault
Damour for useful discussions, and to thank the IH\'{E}S for hospitality during much
of this work. This work has been partially supported by grants from the National
Science Foundation and the Templeton Foundation.

\end{document}